\documentclass[aps, prl, twocolumn, letterpaper]{revtex4-2}

\usepackage{graphicx}
\usepackage{amssymb}
\usepackage{amsmath}
\usepackage{float}
\usepackage{txfonts}
\usepackage{bm}
\usepackage{color}
\usepackage[colorlinks=true, linkcolor=blue, anchorcolor=black, citecolor=blue, filecolor=black, menucolor=black, runcolor=black, urlcolor=blue]{hyperref}
\usepackage{mathrsfs}

\begin{document}

\preprint{AIP/123-QED}

\title{Unsupervised Learning of Topological Non-Abelian Braiding in Non-Hermitian Bands}

\author{Yang Long$^1$}
\email{Corresponding author: yang.long@ntu.edu.sg}
\author{Haoran Xue$^2$}
\author{Baile Zhang$^{1,3}$}
\email{Corresponding author: blzhang@ntu.edu.sg}
\affiliation{%
$^1$Division of Physics and Applied Physics, School of Physical and Mathematical Sciences, Nanyang Technological University, 21 Nanyang Link, Singapore 637371, Singapore \\
$^2$Department of Physics, The Chinese University of Hong Kong, Shatin, Hong Kong SAR, China \\
$^3$Centre for Disruptive Photonic Technologies, Nanyang Technological University, Singapore 637371, Singapore
}%

\date{\today}

\begin{abstract}
The topological classification of energy bands has laid the groundwork for the discovery of various topological phases of matter in recent decades. While this classification has traditionally focused on real-energy bands, recent studies have revealed the intriguing topology of complex-energy, or non-Hermitian bands. For example, the spectral winding of complex-energy bands can from unique topological structures like braids, holding promise for advancing quantum computing. However, discussions of complex-energy braids have been largely limited to the Abelian braid group $\mathbb{B}_2$ for its relative simplicity, while identifying topological non-Abelian braiding is still difficult since it has no universal topological invariant for characterization. 
Here, we present a machine learning algorithm for the unsupervised identification of non-Abelian braiding of multiple complex-energy bands. The consistency with Artin's well-known topological equivalence conditions in braiding is demonstrated. 
Inspired by the results from unsupervised learning, we also introduce a winding matrix as a topological invariant in charactering the braiding topology and unveiling the bulk-edge correspondence of non-Abelian braided non-Hermitian bands. Finally, we extend our approach to identify non-Abelian braiding topology in 2D/3D exceptional semimetals and successfully address the unknotting problem in an unsupervised manner.
\end{abstract}
                    
\maketitle

Braiding is a branch of topology whose history can be traced to ancient time~\cite{Katritch1996}. In modern science, it has played a significant role in the realms of  condensed matter physics~\cite{Wu_2019, Lian2017, Sun2017, Wu2020, Lee2020, Peng2022, Guo2021, Bouhon2020, Patil2022}, non-Abelian anyon~\cite{Nayak2008, Nakamura2020,  Andersen2023, Stern2010},  non-Abelian quantum gates~\cite{AbdumalikovJr2013, Stern2013}, as well as in non-Abelian photonics~\cite{Leach2004, Kedia2013,Pisanty2019,Pisanty2019a, Wang_2021, Zhang2022, Yang2019, Sun2022} and acoustics~\cite{Chen2021, Jiang2021, Zhang2021, You2022}. 
Only very recently has it been realized that the braiding can be used as an essential tool to characterize and classify the topology of complex-energy, or non-Hermitian bands, in parallel with the topological classification of various topological phases of matter such as topological insulators and semimetals that are Hermitian~\cite{Zhang_2023, Wang_2021, Wang2021, Bergholtz_2021, Hu_2021}. 
So far, most discussions about complex-energy bands braiding are applicable only to a two-band Hamiltonian~\cite{Zhang_2023, Wang_2021},  corresponding to the simple Abelian braiding (the braid group $\mathbb{B}_2$) that can be identified with direct visualization.  

Nevertheless, braiding more than two bands can be non-Abelian~\cite{Atiyah1990, Hu_2021, Guo_2023, Li2023} and geometrically complicatd. Firstly, it is difficult to identify with only direct visualization. Secondly, although theoretically the topological characterization can always be accomplished by calculating a certain topological invariant, the selection of the topological invariant in most cases is try and error, due to the lacking of a universal topological invariant. For example, the Alexander polynomial~\cite{Alexander1928} and Jones polynomial~\cite{Jones1985} are widely used to characterize braiding, but two non-equivalent braids can have the same Alexander polynomial or Jones polynomial~\cite{Murasugi2008}. 
Thus, it is desirable to develop an approach, without relying on any topological invariant or any braiding theory, to identify non-Abelian braiding for complex-energy bands. 

Unsupervised learning is a type of machine learning approaches that has rapidly emerged as a powerful tool to uncover latent patterns or underlying principles within raw data, especially where predefined labels are absent~\cite{Carleo2019,Dunjko2018, Mehta2019}. 
Unsupervised learning has found widespread application in the characterization of topological phases of matter~\cite{Rodriguez_Nieva_2019, Scheurer2020, Long_2020}. 
For example, unsupervised learning can be exploited to obtain the topological periodic table without relying on expert knowledge about group theory~\cite{Long_2023}.  
There are some very recent attempts for identifying braiding topology,  including supervised neural network for detecting knotted polymer~\cite{Vandans2020},  machine-learning-guided framework  to find new relations in knot geometry~\cite{Davies2021}, and diffusion map of the simple Abelian braiding~\cite{Yu_2022}. 
However, so far, the non-Abelian braiding in physical systems has not been explored with unsupervised learning. 

\begin{figure*}[tp!]
\centering
\includegraphics[width=\linewidth]{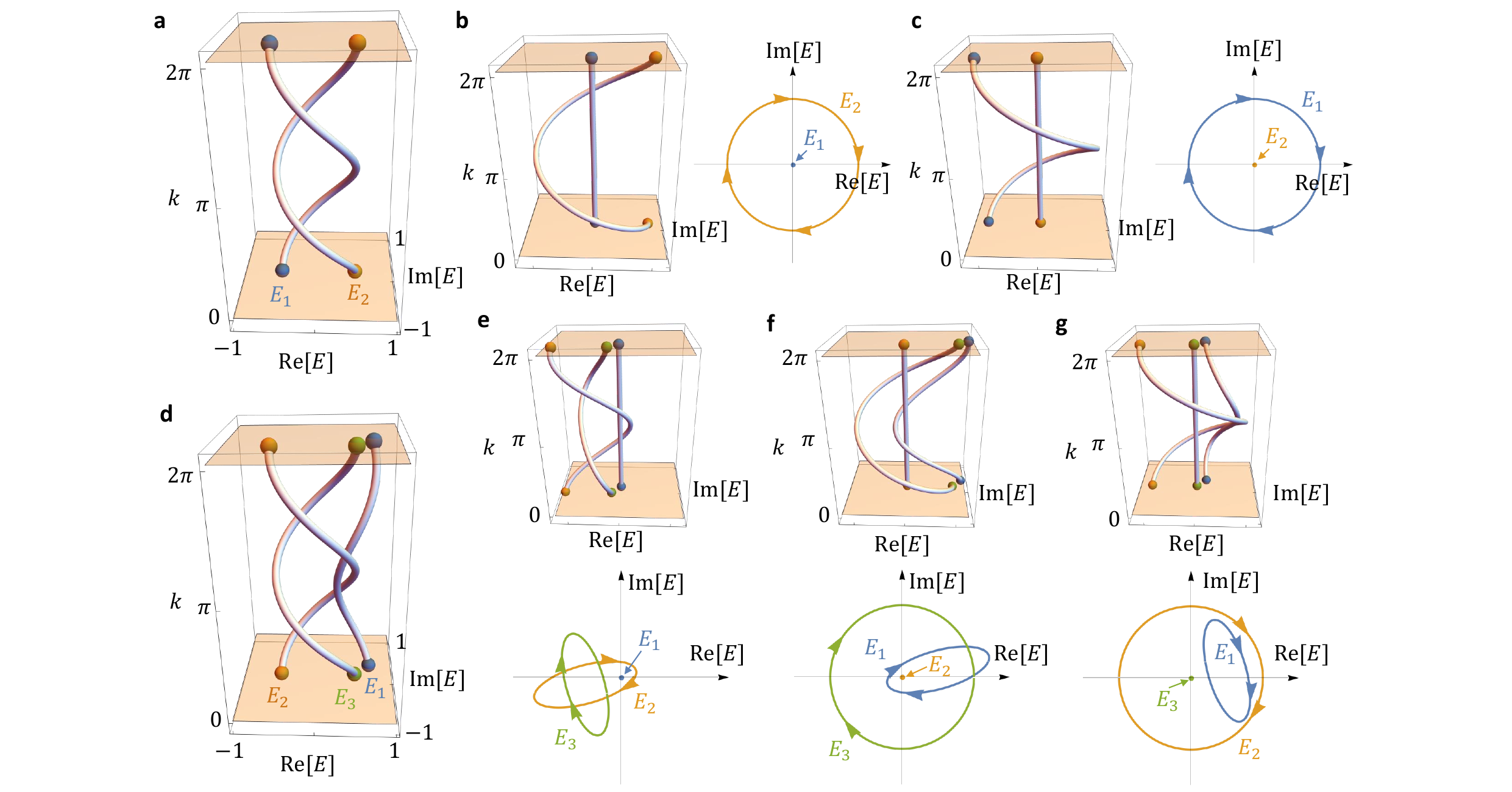}
\caption{\textbf{Non-Abelian braiding topology in non-Hermitian bands}. (a) Abelian braiding between two bands $\{ E_1, E_2 \}$ in a 1D non-Hermitian system.  (b) Topologically equivalent deformation of the braiding in (a) while keeping $E_1$ as a constant (i.e., $0$).  After the deformation, the braiding can be characterized by the winding of $E_2$ with the reference energy as $E_1$. 
(c) Topologically equivalent deformation of the braiding in (a) while keeping $E_2$ as a constant. The braiding can be characterized by the winding of $E_1$ with the reference energy as $E_2$. 
(d) Non-Abelian braiding between three bands $\{ E_1, E_2, E_3\}$ in a 1D non-Hermitian system. 
(e) Topologically equivalent deformation of the braiding in (d) while keeping $E_1$ as a constant.  The braiding can be characterized by the winding of $\{E_2, E_3\}$ with the reference energy as $E_1$. 
(f) Topologically equivalent deformation of the braiding in (d) while keeping $E_2$ as a constant.  The braiding can be characterized by the winding of $\{E_1, E_3\}$ with the reference energy as $E_2$. 
(g) Topologically equivalent deformation of the braiding in (d) while keeping $E_3$ as a constant.  The braiding can be characterized by the winding of $\{E_1, E_2\}$ with the reference energy as $E_3$. 
}
\label{fig:braid_winding}
\end{figure*}

In this Letter, we present an unsupervised learning approach for the identification of topological non-Abelian braiding in non-Hermitian systems. 
We first investigate distinct winding behaviors after performing topologically equivalent deformations of non-Abelian braids in multiple complex-energy bands. 
We introduce a similarity function to quantify the topological distinctions among these non-Abelian braids. 
Utilizing our previously proposed clustering algorithm~\cite{Long_2023}, we can determine the number of topologically distinct braids and group them into clusters. 
The results showcase that our algorithm effectively and accurately identifies topologically distinct non-Abelian braids. 
We unveil that our algorithm can be consistent with the topological equivalence conditions established by E. Artin~\cite{Artin_1947}, confirming our algorithm's reliability. 
Inspired by the clustering results in our algorithm, we introduce a matrix-based topological invariant: the winding matrix, which can not only reflect the non-Abelian feature of topological invariant, but also describe the appearance of non-Hermitian skin effect in non-Abelian braided multi-band system. 
Finally, we extends our algorithm to identify exceptional nodal points/lines described by braiding topology in two/three-dimensional (2D/3D) semimetals and solve the unknotting problem. 

We begin with 1D systems to introduce braiding of bands. 
For 1D non-Hermitian Hamiltonian $H(k)$, we can obtain the energy bands by diagonalizing $H(k)$: $E(k) = {\rm diag}[E_1(k), E_2(k), ..., E_n(k)] = \langle \phi_m(k) | H(k) | \psi_n(k)\rangle
$, where $E_n(k) \in \mathbb{C}$ is the energy of the $n$-th band, $|\psi_n(k)\rangle$ ($|\phi_n(k)\rangle$) is the right (left) eigenvector. 
Since the momentum $k$ is defined in the first Brillouin zone (BZ, $k \in [0,2\pi]$), each energy band, as described by the complex energy $E_n(k)$, can be regarded as a strand of a braid in the (${\rm Re}[E_n]$, ${\rm Im}[E_n]$, $k$)-space. 
Owing to the 1D periodicity, the left and right boundaries of the BZ are equivalent, allowing the closure of braids at BZ boundaries to form knots or links. 
Here, we assume that the complex-energy bands, denoted as ${E_n(k)}$, satisfy the separable band condition, namely, the absence of any crossing point between any two bands. 
According to this condition, two braids are regarded as topologically distinct if the continuous deformation between them necessitates the crossing of at least two bands with one another.

\begin{figure*}[tp!]
\centering
\includegraphics[width=\linewidth]{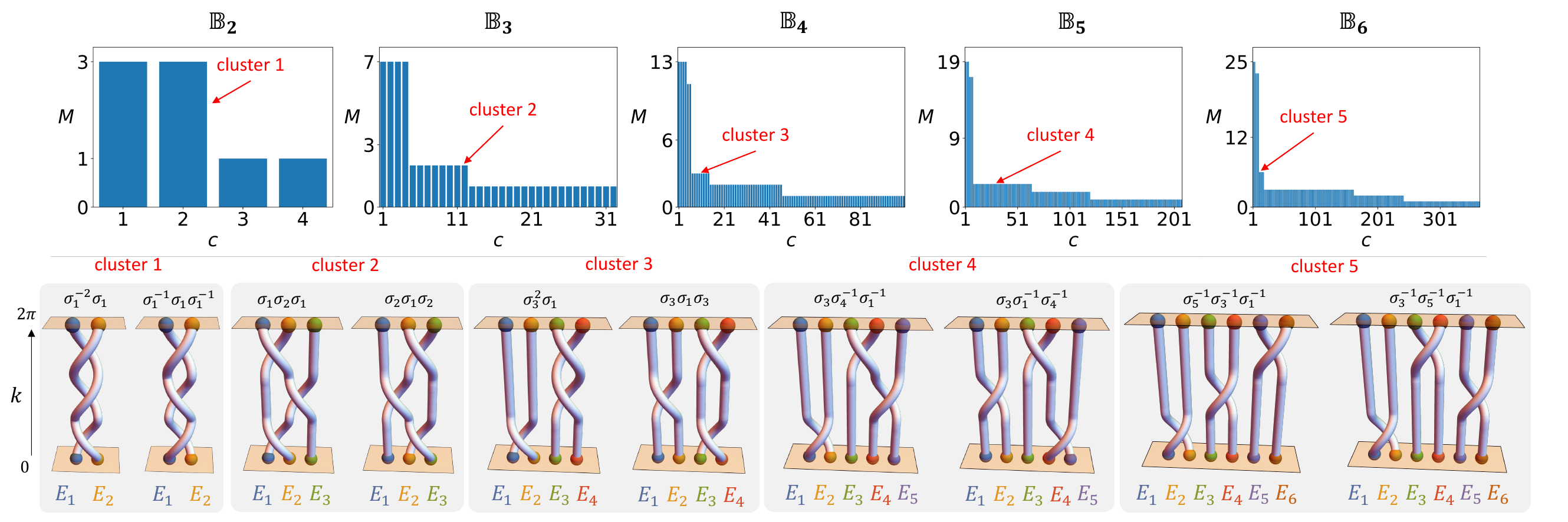}
\caption{\textbf{Unsupervised identification of topological non-Abelian braids}. We generate all combinations by randomly selecting 3 braid operations in different braid groups $\mathbb{B}_2$, $\mathbb{B}_3$, $\mathbb{B}_4$, $\mathbb{B}_5$ and $\mathbb{B}_6$.  After performing unsupervised learning algorithm, we plot two braids in typical clusters: (1) cluster 1 in $\mathbb{B}_2$ ($c=2$); (2) cluster 2 in $\mathbb{B}_3$ ($c=12$);  (3) cluster 3 in $\mathbb{B}_4$ ($c=12$); (4) cluster 4 in $\mathbb{B}_5$ ($c=34$); (5) cluster 5 in $\mathbb{B}_6$ ($c=16$).  
From the plotted braids in the different clusters (i.e., in the same phases), we can see that our algorithm can be consistent with the Artin's topological equivalent conditions in the braiding, which guarantees the success of our algorithm in classifying the topological phases among braids. More details can be found in the Supplementary Material~\cite{Note1}.  }
\label{fig:classify_braid}
\end{figure*}

We firstly investigate the topology of braids in $\mathbb{B}_2$, which is isomorphic to $\mathbb{Z}$. 
We take the braiding in Fig.~\ref{fig:braid_winding}(a) as an example. 
The braiding topology between two bands, as illustrated in Fig.\ref{fig:braid_winding}(a), can be geometrically conceptualized as one band winding around the other~\cite{Zhong2018}. 
In Fig.\ref{fig:braid_winding}(b), the braid can undergo a continuous transformation such that $E_1$ becomes a constant. This allows us to represent the topology by the winding behavior of $E_2$ around $E_1$. 
Alternatively, we can also fix $E_2$ as a constant to obtain a  geometrically different representation, but  the winding topology maintains the sme, as shown in Fig.~\ref{fig:braid_winding}(c). 
Thus, the topological properties in the braiding between two bands $\{E_1, E_2\}$ can be described by the winding of $E_1$ around $E_2$ or reverse~\cite{Note1},  similar to the exchange statistics of anyons in topological field theory~\cite{arxiv.2210.02530}. 
We can extend this insight to encompass non-Abelian braiding involving three bands, namely ${E_1, E_2, E_3}$, as depicted in Fig.\ref{fig:braid_winding}(d). 
In Fig.\ref{fig:braid_winding}(e), when we keep $E_1$ as a constant, we can plot the winding behaviors of $E_2$ and $E_3$ around $E_1$. 
Similarly, when we fix $E_2$ or $E_3$ as a constant, we can determine the winding behaviors of the remaining bands, as plotted in Fig.~\ref{fig:braid_winding}(e,f). 
In contrast to the Abelian braiding of two bands, the winding behaviors in a non-Abelian braid of multiple bands can differ when we fix different bands as a reference energy. 
This observation shows that topological properties in a non-Abelian braid can be associated with different winding behaviors with respect to different bands. 

To describe the winding behaviors, we introduce a Hermitian Hamiltonian $\mathcal{H}_{mn}$ under chiral symmetry, which is related to the winding of $E_m$ around $E_n$~\cite{Note1}:
\begin{equation}
\mathcal{H}_{mn} = \begin{pmatrix}
0 & E_m(k) - E_n(k) \\
E^*_m(k) - E^*_n(k) & 0 
\end{pmatrix}
\label{eq:Hamiltonian_chiral}
\end{equation}
Here $\mathcal{H}_{mn}$ satisfies the chiral symmetry $\tau_z \mathcal{H}_{mn} \tau_z = - \mathcal{H}_{mn}$, where $\tau_z$ is the Pauli matrix. 
According to the previous analysis on the point-gap topology~\cite{Kawabata2019}, the band topology of $\mathcal{H}_{mn}$ can have a one-to-one correspondence with the point-gap topology of $E_m$ when $E_n$ is the reference energy. 
Note that $\mathcal{H}_{nm}$ shares the same topology with $\mathcal{H}_{mn}$ since $\mathcal{H}_{nm}=-\mathcal{H}_{mn}$.
To cover all winding properties in an $N$-band non-Hermitian system, a collection of in total $N(N-1)/2$ Hamiltonians, denoted as $\{\mathcal{H}_{mn}|m\neq n\}$, are calculated. 
To judge on the topological distinction between braids, we propose the following similarity function $\mathcal{K}_{ij}$ between two samples $H_i$ and $H_j$~\cite{Note1}:
\begin{equation}
\mathcal{K}_{ij} = \Pi_{k \in {\rm BZ}} \Pi_{m\neq n} \left(1- e^{-\frac{|\det[\mathcal{Q}^{(i)}_{mn}(k) + \mathcal{Q}^{(j)}_{mn}(k)]|^2}{\epsilon^2}}\right)
\end{equation}
Here $\mathcal{Q}^{(i)}_{mn}(k)$ is the flatten Hamiltonian of $\mathcal{H}_{mn}$ formed by the complex-energy bands in $H_i$, BZ denotes the BZ of $\mathcal{H}_{mn}$~\cite{Note1}. 
$\mathcal{Q}_{mn}(k) = 1-2\mathcal{P}_{mn}(k)$ and $\mathcal{P}_{mn}(k) = \sum_{n' \in {\rm occ}} |\varphi_{n'}(k)\rangle \langle \varphi_{n'}(k)|$, where $|\varphi_{n'}(k)\rangle$ is the eigenvector of $\mathcal{H}_{mn}$: $\mathcal{H}_{mn} |\varphi_{n'}(k)\rangle = \mathcal{E}_{n'}(k) |\varphi_{n'}(k)\rangle$, and ${\rm occ} = \{n' | \mathcal{E}_{n'}(k) < 0\}$. 
In calculating similarities, random chiral-symmetry-preserved perturbations are added to improve the numerical accuracy and test topological robustness~\cite{Long_2023, Note1}. 
More details can be found in Supplementary Material~\footnote{See Supplementary for more details, including the details of unsupervised learning, generation scheme of samples, winding matrix, 2D/3D exceptional semimetal and unknotting problem}.

We exploit our previously proposed clustering algorithm in Ref.~\cite{Long_2023} to detect the number of phases and identify the braids in the Hamiltonian samples $\{ H_i \}$. Here, we review the algorithm briefly. 
There are a set $\mathcal{S}$ and a list $\{M_c\}$ in our algorithm: $\mathcal{S} = \{H_{p_c}\}$ is a set of samples that are mutually different (i.e., $\mathcal{K}_{p_c p_{c'}}<1/2$, $\forall H_{p_c}, H_{p_{c'}}\in \mathcal{S}$) and $M_c$ denotes the number of samples that are topologically equivalent to $H_{p_c}$, $\{M_c|c=1,2,...N_c\}$. 
The algorithm contains the following two steps: (1) Add the first sample $H_1$ into $\mathcal{S}$ since the initial $\mathcal{S} = \emptyset$. Then, $\mathcal{S}=\{H_1\}$, $p_1=1$, $M_1 = 1$ and $N_c=1$. 
(2) Compare the following sample $H_j$ with the samples in $\mathcal{S}$: if $H_j$ is topologically equivalent to $H_{p_c}$, i.e., $H_{p_c}\in \mathcal{S}$ and $\mathcal{K}_{j,p_c}> \kappa_c$,  then $M_c := M_c + 1$. Otherwise, if none of the samples in $\mathcal{S}$ is topologically equivalent to $H_j$, we add  $H_j$ into  $\mathcal{S}$, $M_{N_c+1} = 1$, $p_{N_c+1}=j$ and $N_c := N_c + 1$. 
After calculating all samples in $\{ H_i \}$, we can obtain: $N_c$ denotes the number of topologically distinct phases, and $\{M_c\}$ denotes the number of samples that have the same phase as $H_{p_c}$~\cite{Note1}. We use $c$ to denote the label of topologically distinct phases.  

\textit{Identify the topology in non-Abelian braiding}. 
Braiding can be represented using Artin's braid words, denoted as $\{\sigma_i\}$~\cite{Artin_1947, Atiyah1990}. The braid operation $\sigma_i$ represents the interchange of the lower endpoints of the $i$-th and ($i+1$)-th strands while maintaining the upper endpoints fixed, with the $i$-th strand crossing above the ($i+1$)-th strand. Conversely, if the $i$-th strand crosses beneath the ($i+1$)-th strand, it is denoted as $\sigma_i^{-1}$. 
In Fig.~\ref{fig:classify_braid}, we generate random examples with three braid operations in non-Hermitian systems with different numbers of bands, belonging to different braid groups. For bands in $\mathbb{B}_N$, we randomly select three braid operations from the set ${\sigma_1, \sigma_2,..., \sigma_{N-1}, \sigma^{-1}_1, \sigma^{-1}_2,..., \sigma_{N-1}^{-1}}$, allowing choosing a braiding operation multiple times. 
The number of possible samples for selecting three braid operations in $\mathbb{B}_N$ is $(2(N-1))^3$. 

The classification results for randomly generated braids belonging to $\mathbb{B}_2$, $\mathbb{B}_3$, $\mathbb{B}_4$, and $\mathbb{B}_5$ are presented in Fig.~\ref{fig:classify_braid}. 
Here we choose five distinct clusters labeled as $1$ to $5$, and for each cluster, we showcase two representative samples. 
It is evident from the braids within these clusters that our algorithm effectively classifies operations sharing the same topological phase. 
For example, the topological characteristics of the Abelian group $\mathbb{B}_2$ can be described by the numerical difference between $\sigma_1$ and $\sigma_1^{-1}$ in the braid words, which is isomorphic to $\mathbb{Z}$. 
In cluster 1, the two braids are identical as they both contain 2 instances of $\sigma_1^{-1}$ and 1 instance of $\sigma_1$. 
However, when dealing with $\mathbb{B}_N$ ($N>2$), topological properties are non-Abelian in braid operations. 
The two braids within each of clusters 2 to 5 can undergo continuous deformations into each another, showing that our algorithm aligns with theoretical predictions in non-Abelian braiding. 
Furthermore, our algorithm accommodates topological classification between braids of varying lengths~\cite{Note1}. 

The effectiveness of our algorithm in identifying braiding topology can be verified by its consistency with the well-known Artin's topological equivalence conditions. 
According to the braid group theory~\cite{Artin_1947}, the following braid operations are considered topologically equivalent: $\sigma_i \sigma_j = \sigma_j \sigma_i$ ($|i-j|\leq 2$) and $\sigma_i \sigma_{i+1} \sigma_i = \sigma_{i+1} \sigma_i \sigma_{i+1}$. 
As illustrated in Fig.~\ref{fig:classify_braid}, it is apparent that the braids within the clusters associated with $\mathbb{B}_N$ ($N>2$) conform to Artin's topological equivalence conditions. 
For example, cluster 2 exhibits braids like $\sigma_1 \sigma_2 \sigma_1$ and $\sigma_2 \sigma_1 \sigma_2$, and cluster 3 contains braids such as $\sigma_3^2 \sigma_1$ and $\sigma_3 \sigma_1 \sigma_3$, all satisfying these conditions. 
To validate our algorithm's consistency with Artin's topological equivalence conditions, we have conducted various tests~\cite{Note1}, including assessing random sequences of braid operations in the braid word, appending random braid words, and performing combinations of these tests. 

\begin{figure}[tp!]
\centering
\includegraphics[width=\linewidth]{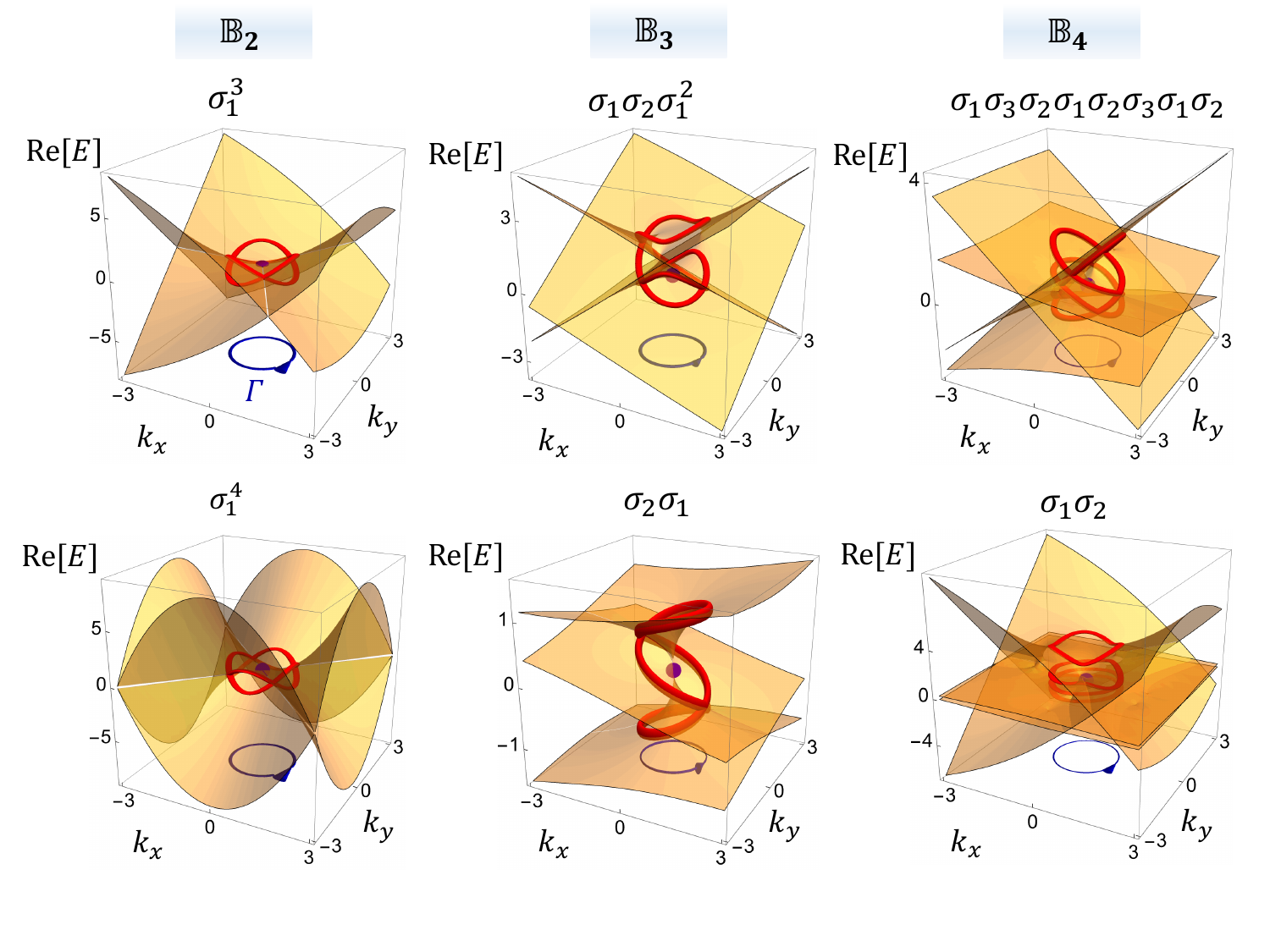}
\caption{\textbf{2D expcetional topological semimetal with non-Abelian braiding topology}. 
The topology around the central exceptional  nodal point is described by the  band topology on the sphere $S^1$ denoted as $\Gamma$ (shown in an arrowed blue circle). 
The bands on $S^1$ (denoted as the red lines) possess the braiding topology, which is described as the braid words that are shown on the top of each figure. 
}
\label{fig:2D_semimetal}
\end{figure}

Based on the numerical results presented above, it is apparent that the braiding topology can be characterized with the winding between bands around one another. 
For an $N$-band non-Hermitian system ($N \geq 2$) satisfying the separable band condition, we introduce a winding matrix $w$ to serve as a topological invariant for the braiding topology. The matrix element $w_{mn}$ is~\cite{Note1, Shen2018}:
\begin{equation}
w_{mn} = \frac{1}{2\pi i}\int \partial_k \ln [E_m(k) -E_n(k)] dk
\label{eq:winding_matrix}
\end{equation}
Clearly, $w_{mn}=w_{nm}$. Generally, $w_{mn}$ is not quantized and $w$ can be distinct even for two topologically equivalent braids. 
However, after we represent the complex-energy bandstructure into a braid diagram, we can obtain the quantized winding matrix $w$~\cite{Note1}. 
The quantized winding matrix serves as a topological invariant of the non-Abelian topology. 
$w_{mn}$ denotes how many times the $m$-th band winds around the $n$-th band. 
For two bands, $N=2$, $w= \frac{v}{2} \tau_x$ ($v \in \mathbb{Z}$, and $\tau_x$ is a Pauli matrix), which has a one-to-one correspondence to an integer number $v$~\cite{Note1}.  
Thus, for $N=2$, $w$ is Abelian, corresponding to the Abelian property of $\mathbb{B}_2$ (i.e., $\mathbb{B}_2$ is isomorphic to $\mathbb{Z}$). 
For $N>2$, $w$ cannot be represented by only a single integer. 
The non-commutative nature of $w$ means that $w$ becomes non-Abelian, namely, $w^{(1)} w^{(2)} \neq w^{(2)} w^{(1)} $, where $w^{(m)}$ denotes the winding matrix of $m$-th sample, being consistent with the fact that $\mathbb{B}_N$ ($N>2$) is non-Abelian. 

The winding matrix not only  characterizes the non-Abelian topology in braiding, but also establishes a bulk-edge correspondence. 
The presence of braiding topology can lead to the manifestation of the non-Hermitian skin effect, where an extensive number of bulk modes get localized at an end of a sample, forming so-called skin modes. 
The difference in the number of oppositely localized skin modes is directly linked to the winding matrix $w$: $N_s = N_l - N_r = \sum_{m,n} w_{mn}$, where $N_{l,r}$ denotes the number of skin modes that localize at the left or right end of a finite lattice ~\cite{Note1}. 
Note that $N_s = 0$ does not mean the disappearance of non-Hermitian skin effect, because the skin modes that localize at the opposite ends can appear in pair, akin to the bipolar non-Hermitian skin effect~\cite{Song_2019} or $\mathbb{Z}_2$ skin effect~\cite{Okuma2020}. 
More details can be found in Supplementary Material~\cite{Note1}.

\begin{figure}[tp!]
\centering
\includegraphics[width=\linewidth]{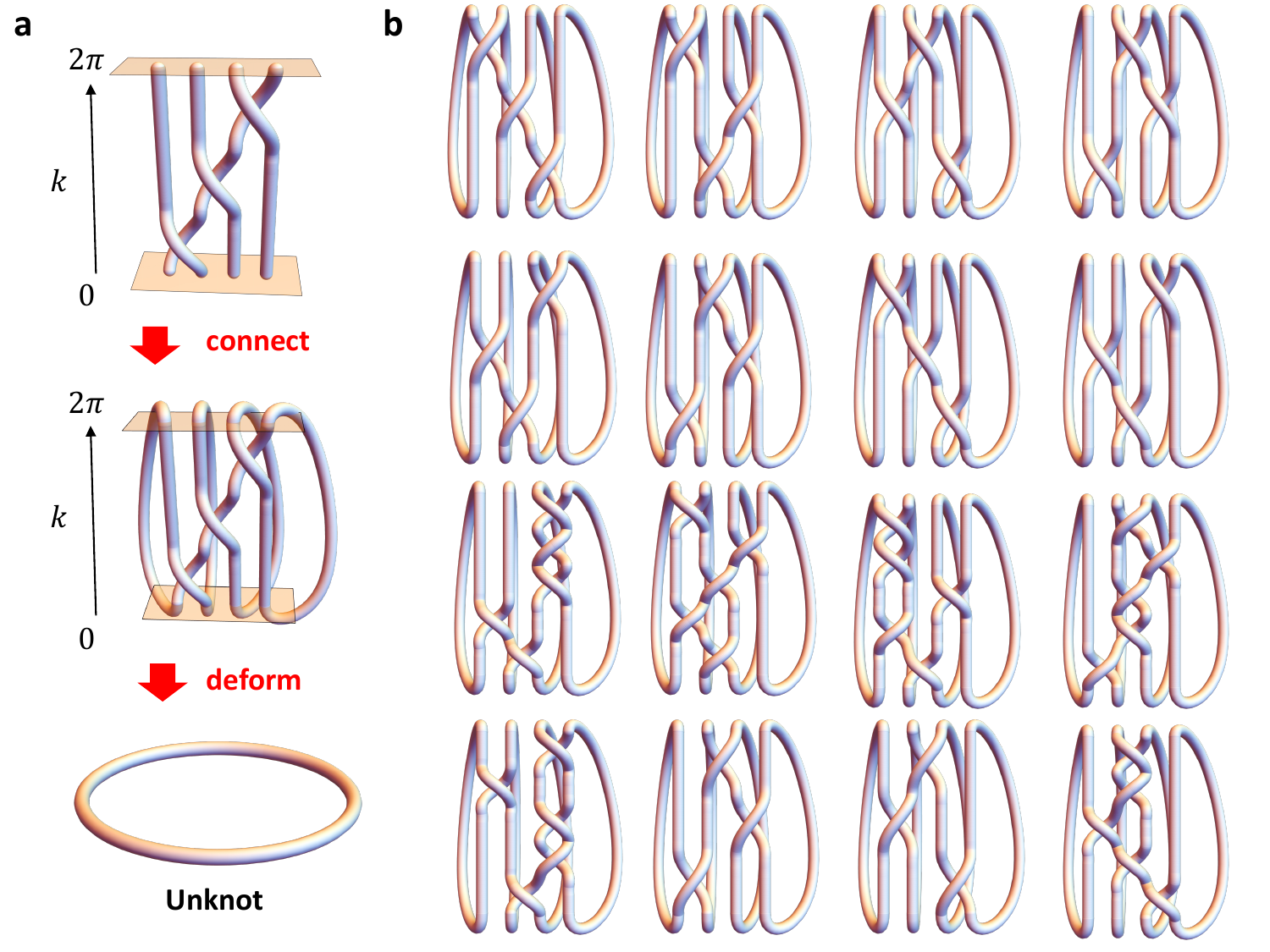}
\caption{\textbf{Solving the unknotted problem unsupervisedly}. (a) The braids can be closed due to the periodicity of BZ. Here, we show an unknotted bandstructure ($\sigma_3\sigma_2\sigma_1$). After connecting the bands, we can continuously deform the resultant structure into an unknot. 
(b) Some unknots identified by our algorithm.  We firstly generate all braid words of unknots for $N$ bands and form a group $\mathcal{S}$. Here, we set $N=4$. Then, we randomly generate the braids that belong to $\mathbb{B}_4$. We then perform the similarity function to measure the difference between the generated samples and the samples in $\mathcal{S}$. The generated braid that is in the same phase as any in $\mathcal{S}$ is an unknot.}
\label{fig:unknot}
\end{figure}

\textit{Non-Hermitian topological semimetal with non-Abelian braiding}. 
We can extend the 1D braiding topology  to the braiding topology within 2D/3D non-Hermitian topological semimetals. 
According to topological classification theory, 2D gapless topological systems can be linked to 1D topological systems through a mapping relationship~\cite{Kawabata2019, Kawabata_2019}. 
For example, by employing the transformation $\cos(k) \rightarrow k_x$ and $\sin(k) \rightarrow k_y$, we can construct a 2D gapless system with an exceptional point located at the origin $(0,0)$. 
In Fig.\ref{fig:2D_semimetal}, we present various cases of 2D exceptional points featuring braiding topology, ranging from $\mathbb{B}_2$ to $\mathbb{B}_4$. 
On the sphere $S^{1}$ described by the circle $\Gamma$ (the blue arrowed line in Fig.\ref{fig:2D_semimetal}), we can see the braiding of bands (denoted as red lines in Fig.\ref{fig:2D_semimetal}), which belong to $\mathbb{B}_3$ and $\mathbb{B}_4$, respectively. 
Note that the braiding topology of the 2D exceptional topological semimetal discussed here is not the knot or link structure formed by the exceptional nodal lines~\cite{Carlstr_m_2019, Bi_2017, Bergholtz_2021}. 
The  winding matrix $w$ of Eq.~\ref{eq:winding_matrix} can serve as a topological invariant for characterizing exceptional nodal points or lines, similar to the use of non-Abelian quaternion charge in describing nodal line semimetals~\cite{Wu_2019}. 
We can also extend the 1D braiding topology of non-Hermitian bands to the braiding topology in 3D non-Hermitian exceptional nodal lines, in which the exceptional nodal line undergoes non-Abelian braiding of bands on the manifold sphere $S^1$ that encloses it~\cite{Note1}. 

\textit{Solving the unknotting problem unsupervisedly}. 
In addition to identifying the braids formed by non-Hermitian bands, our algorithm can also be applied to solve certain geometrical braiding problems. 
Here, we take the unknotting problem as an example. The unknotting problem is about detecting whether a braid can be unknotted. 
For instance, as illustrated in Fig.~\ref{fig:unknot}(a), we can connect all bands due to the periodicity of the BZ and subsequently deform the resulting closed braid into an unknot. 
While there are theoretical approaches to solving this problem, such as Haken's algorithm~\cite{Haken1961} or Khovanov homology~\cite{Kronheimer2011}, they generally involve intricate theoretical processes and tend to be time consuming. 
Besides traditional theoretical approaches, some supervised-learning-based approaches are proposed recently~\cite{Gukov_2021, Craven2023}. 
Here, we demonstrate how our unsupervised algorithm can be utilized to solve the unknotting problem. 

We can determine whether a given braid is an unknot by computing its similarity to all the generated unknots. If a briad is topologically equivalent to any of the generated unknots, it can be classified as an unknot. 
To illustrate our approach for solving the unknotting problem, we  assume that we have the braid word representing the target closed braid, which is denoted as $\sigma_{a_1}\sigma_{a_2} ...\sigma_{a_{n_c}}$. 
As an example, we show how to find unknots in a system with 4 bands ($N=4$). 
Firstly,  we generate a set $\mathcal{S}_0$ that contains all possible unknots for $N=4$~\cite{Note1}. 
For instance, $\mathcal{S}_0$ includes braids like $\sigma_{N-1}\sigma_{N-2}...\sigma_2\sigma_1$. 
Then, we randomly generate Hamiltonians $\{H_i\}$ with various braids by selecting and combining braid operations in ${\sigma_1, \sigma_2..., \sigma_{N-1}, \sigma_1^{-1}, \sigma_2^{-1},...,\sigma_{N-1}^{-1} }$. 
We calculate the similarity between each generated braid sample $H_i$ and all members of the set $\mathcal{S}_0$. 
If $H_i$ is topologically equivalent to any braid in $\mathcal{S}_0$, it can be classified as an unknot. 
As depicted in Fig.~\ref{fig:unknot}(b), we demonstrate 16 unknots that are identified among the randomly generated samples.

In summary, we have developed an unsupervised learning algorithm for the topological classification of non-Abelian braiding in non-Hermitian bands. 
Our research not only deepens the current understanding of non-Hermitian topological physics, but also opens new avenues for the application of machine learning in this field, including potential applications in topological sensors~\cite{Hodaei2017, Chen2017, Li2023a, Xu2016}, topological lasers~\cite{Harari2018, Bandres2018, Zeng2020, Zhu2022}, and topological many-body systems~\cite{Fruchart2021, Matsumoto2020}.


\begin{acknowledgments}
This research is supported by Singapore National Research Fundation Competitive Research Program under Grant no. NRF-CRP23-2019-0007, and Singapore Ministry of Education Academic Research Fund Tier 2 under Grant No. MOE-T2EP50123-0007. 
H.X. acknowledges support from the startup fund of The Chinese University of Hong Kong. 
Y.L. gratefully acknowledges the support of the Eric and Wendy Schmidt AI in Science Postdoctoral Fellowship, a Schmidt Futures program. 
\end{acknowledgments}

All the code necessary for reproducing our results is publicly available at https://github.com/XXX. 
An archived version has also been deposited in the Zenodo database https://doi.org/XXX.

\bibliography{references}

\end{document}